\documentclass[final,1p,times]{elsarticle}

\usepackage{amsmath,amssymb,bm}
\usepackage{comment}

\begin{document}

\begin{flushright}
KEK-TH-2055 and J-PARK-TH-0130
\end{flushright}

\begin{frontmatter}



\title{The role of electron orbital angular momentum in the 
Aharonov-Bohm effect revisited}


\author[label1,label2]{Masashi Wakamatsu\corref{cor1}}
\ead{wakamatu@post.kek.jp}
\author[label1]{Yoshio~Kitadono}
\author[label1]{Liping Zou}
\author[label1]{Pengming Zhang}

\cortext[cor1]{corresponding author}

\address[label1]{Institute of Modern Physics, Chinese Academy of Sciences,\\
Lanzhou, People's Republic of China, 730000}
\address[label2]{KEK Theory Center, Institute of Particle and Nuclear Studies,\\
High Energy Accelerator
Research Organization (KEK),\\
1-1, Oho, Tsukuba, Ibaraki 305-0801, Japan}

\begin{abstract}
This is a brief review on the theoretical interpretation of the 
Aharonov-Bohm effect, which also contains our new insight into
the problem. A particular emphasis is put on the unique role
of electron orbital angular momentum, especially viewed from the 
novel concept of the physical component of the gauge field, which
has been extensively discussed in the context of the nucleon spin 
decomposition problem as well as the photon angular momentum
decomposition problem.
Practically, we concentrate on the frequently discussed idealized
setting of the Aharonov-Bohm effect, i.e. the interference phenomenon
of the electron beam passing around the infinitely-long solenoid.
One of the most puzzling observations in this
Aharonov-Bohm solenoid effect is that the pure-gauge
potential outside the solenoid appears to carry non-zero orbital angular
momentum. Through the process of tracing its dynamical origin,
we try to answer several fundamental questions of the
Aharonov-Bohm effect, which includes the question about the
reality of the electromagnetic potential, the gauge-invariance
issue, and the non-locality interpretation, etc. 
\end{abstract}


\end{frontmatter}



\section{Introduction}
\label{Section:s1}

Now, there is no doubt about the existence of the
Aharonov-Bohm effect (AB-effect) \cite{AB1959}, also called the
Ehrenberg-Siday-Aharonov-Bohm effect \cite{ES1949}, 
especially after Tonomura et al.'s experiments with use of a
toroidal magnet field confined by a superconductor was
performed \cite{Tonomura1986},\cite{Osakabe1986}.
Concerning the theoretical interpretation of the observed AB-phase,
however, the most commonly discussed case is the interference
phenomenon of the electron beam passing around an infinitely-long
solenoid, sometimes called the Aharonov-Bohm solenoid effect.
Although this setting of the problem is believed to contain the
essence of the AB-effect, an extreme idealization sometimes causes
controversies and it appears to be a reason why many 
seemingly-conflicting interpretations of the AB-effect coexist.
(Earlier reviews of the AB-effect can, for instance, be found 
in \cite{Peshkin1981}, \cite{GP1985}, \cite{PT1989}.)

\vspace{2mm}
Let us start with basic theoretical questions about
the AB-effect as : 

\begin{itemize}
\item Is it a purely quantum-mechanical effect with no classical analogue ?

\item Does it show the reality of electromagnetic potentials ?
Namely, are they no longer mathematical tools with more fundamental 
significance than the electromagnetic fields ?
\end{itemize}

The answer to the first question is generally believed to be ``Yes'',
although there are some attempts to explain the AB-effect by the
action of the classical force \cite{Boyer2000A},\cite{Boyer2000B}.
The 2nd question is not so easy to answer, and debate is still continuing.
In fact, a delicate point of the AB-effect is well known.
The AB-phase is related to the closed circular
integral of the vector potential outside the solenoid not to the
vector potential itself.
Then, although the AB-effect certainly comes from the non-zero
vector potential in the magnetic field free region, 
the closed circular integral of the vector potential
and consequently the observable AB-phase depends only on the
net magnetic flux inside the solenoid. (This fact is believed
to be important for gauge-invariance of the AB-effect.)
This means that one can explain the AB-effect with use of the
electromagnetic field only, but the effect depends on the field
values in a region from which the test particle is excluded.
On the other hand, if one genuinely accepts to use the electromagnetic
potential, one can say that the AB-effect depends only on the
potential in the region where the electron is allowed to move.
One must therefore choose either of the following two options :  

\begin{itemize}
\item abandon the principle of locality, or the ``action-through-medium principle''

\item accept that the electromagnetic potential offers a more complete 
description of the electromagnetism than the electric and magnetic
fields at least in quantum mechanics.

\end{itemize}

``Non-locality or reality of electromagnetic potential ?'' 
That is the question one is faced 
with \cite{Vaidman2012},\cite{APP1969},\cite{AK2004},\cite{Tiwari2017}. 
One caution here is that,
if one takes the standpoint that favors the reality of the
electromagnetic potential, one must make one's attitude clear to
the standard belief that its observability contradicts
the so-called " gauge principle", because it is a wide-spread belief that
the electromagnetic potential is not a gauge-invariant quantity.

One of the main purposes of the present paper is to show that the above
two options are not so conflicting claims as is widely believed.
First, in sect.2, we try to establish the concept of physical component
of the electromagnetic field. This includes the standard transverse-longitudinal
decomposition of the vector potential based on the Helmholtz theorem.
According to this theorem, the transverse part of the vector potential
is uniquely fixed and it is expressed in terms of the magnetic field
distribution, which especially means that the transverse component
of the vector potential is gauge-invariant. The resultant transverse
vector potential turns out to be nothing but the vector potential
in the symmetric gauge or in the Coulomb gauge. After pointing out
an intimate connection between the Helmholtz theorem and the
Coulomb gauge choice, we discuss more general Helmholtz theorem called
the causal Helmholtz theorem, which is completely consistent with
the causality and/or the relativity. 
As we shall show, since the magnetic field in the AB-phase
measurement is static, the causal Helmholtz theorem simply
reduces to the standard Helmholtz theorem. Thus, although the
transverse or physical component of the vector potential obtained from
the Helmholtz theorem is expressed with the magnetic field at
the remote place, this non-local feature should not be taken as
violating causality or relativity. We shall argue that the reality interpretation
of the vector potential and the non-locality interpretation of the
AB-effect is not a contradictory idea in this sense.

Next, in sect. 3, we focus on the role of electron orbital angular
momentum in the Aharonov-Bohm solenoid effect, which is another
main objective of the present paper.
Since the vector potential outside the infinitely-long solenoid
takes a pure-gauge configuration, there is no magnetic field
outside the solenoid. 
A curious observation, which was emphasized by Tiwari in a recent
paper \cite{Tiwari2017}, is that this pure-gauge potential appears to carry
non-zero orbital angular momentum. We stress that this peculiar
feature of the pure-gauge configuration is inseparably connected
with the essence of the Aharonov-Bohm solenoid effect. 
What is the ultimate dynamical
origin of non-zero orbital angular momentum associated with 
the pure-gauge potential ?
We compare two totally different explanations based on
totally different local forces, thereby revealing a very singular nature
of the idealized setting of the AB-phase shift measurement, which 
comes from the assumption of infinitely long solenoid.
We also try to clarify the origin of this non-zero angular momentum
from the standpoint of general electromagnetic theory of
angular momentum. 

Finally, in sect.4, we summarize our understanding of the AB-effect
obtained from the present analysis.

\section{Concept of physical component of the photon field}
\label{Section:s2}

The frequently used idealized setting of the AB-effect is an infinitely
long solenoid with some radius $R$. The confined magnetic field inside
this solenoid is expressed as
\begin{equation}
 \bm{B} (\bm{x}) \ = \ B_0 \,\theta (R - r) \,
 \bm{e}_z , \label{B_inf_solenoid}
\end{equation}
where $r = |\bm{x}|$ is the distance from the center axis of the solenoid,
while $\bm{e}_z$ is a unit vector along the $z$-axis.
The corresponding vector potential in the
so-called symmetric gauge (or the Coulomb gauge) is given by
\begin{equation}
 \bm{A} (\bm{x}) \ = \ \left\{
 \begin{array}{ll}
 \frac{1}{2} \,B_0 \,r \,\bm{e}_\phi \ \ & \  
 (r < R) , \\
 \frac{1}{2} \,B_0 \,\frac{R^2}{r} \,\bm{e}_\phi
 \ \ & \ 
 (r \geq R) . \\
 \end{array} \right.
\end{equation}
As is well-known, this naturally explains the AB-phase
\begin{equation}
  \phi_{AB} \ = \ e \,\oint_{|\bm{x}| = r \,(\,> R)} \,
 \bm{A} (\bm{x}) \cdot d \bm{x}
 \ = \ e \,\Phi ,
\end{equation}
where $\Phi \equiv \pi \,R^2 \,B_0$ is the total magnetic flux
through the solenoid.
(In this paper, the natural unit $\hbar = c = 1$ is used, while the electron 
charge is denoted as $e$ with $e = - \,|e|$.)
However, we know that the vector potential is not a gauge-invariant quantity. 
In fact, consider a gauge potential obtained by a sample
gauge transformation from the symmetric gauge as
\begin{equation}
 \bm{A}^\prime \ = \ \bm{A} + \nabla \chi \ \ \ 
 \mbox{\tt with} \ \ \ 
 \chi \ = \ \frac{1}{2} \,B_0 \,
  x \,y \ =  \ 
 \frac{1}{4} \,B_0 \,r^2 \,\sin 2 \,\phi .
\end{equation}
One finds that the new gauge potential is given by
\begin{equation}
 \bm{A}^\prime \ = \ \left\{
 \begin{array}{ll}
 B_0 \, x \,\bm{e}_y \ & \ (r < R) , \\
 \frac{1}{2} \,B_0 \,r \,\left( \cos 2 \,\phi 
 + \frac{R^2}{r^2} \right) \,\bm{e}_\phi \ + \ 
 \frac{1}{2} \,B_0 \,r \,\sin 2 \phi \,\bm{e}_r 
 \ & \  
 (r \geq R) . \\
 \end{array} \right.
\end{equation}
One may notice that the above vector potential inside the solenoid
is an analog of the 2nd Landau gauge, which appears in the Landau
problem, although the vector potential outside the solenoid
takes a little more complicated form.
By this reason, we hereafter call the above gauge choice the
generalized 2nd Landau gauge. Anyhow, one just confirms that the
vector potential is in fact gauge-variant quantity.
Nonetheless, the following identity follows : 
\begin{eqnarray}
 &\,& e \,\oint_{|\bm{x}| \,= \,r \,> \,R} \,
 \bm{A}^\prime (\bm{x}) \cdot d \bm{x} \ = \ 
 e \,\int_0^{2 \,\pi} \,A^\prime_\phi \,\,r \,d \phi \nonumber \\
 &=& 
 e \,\int_0^{2 \,\pi} \,\frac{1}{2} \,B_0 \,
 \left( R^2 + r^2 \,\cos 2 \phi \right) \,d \phi 
 \ = \ e \,\pi \,R^2 \,B_0 \ = \ e \,\Phi \ = \ \phi_{AB} .
\end{eqnarray}
One therefore confirms the well-known fact that the vector potential
is generally gauge-choice dependent, but its closed circular integral, or
the AB-phase, is gauge-invariant.

\vspace{2mm}
Several question naturally arises.

\begin{itemize}
\item There are in principle infinitely many gauge choices.
Still, can one say that only one particular choice of $\bm{A}$ is physical ?

\item What would be the relation between a particular gauge choice and a particular 
path choice in the gauge-invariant formulation of the electrodynamics  
a la DeWitt \cite{DeWitt1962} ? (This question was recently addressed
in our study on the issue of gauge choice in the 
Landau problem \cite{WKZ2018}.)

\end{itemize} 

We emphasize that the second is also a relevant question for our present study.
In fact, remember that DeWitt invented his gauge-invariant formulation of the 
electrodynamics with a special intention of denying the physical reality of the gauge
potential \cite{DeWitt1962}.

\vspace{2mm}
DeWitt's gauge-invariant (but path-dependent ) vector potential
$\tilde{A}_\mu (x)$ is defined as
\begin{equation}
  \tilde{A}_\mu (x) \ \equiv \ A_\mu (x) - 
 \partial_\mu \, \Lambda (x) , \label{A_DeWitt_1}
\end{equation}
with
\begin{equation}
 \Lambda (x) \ \equiv \ \int_0^1 \,A_\sigma (z) \,
 \frac{\partial z^\sigma}{\partial \xi} \,d \xi . \label{A_DeWitt_2}
\end{equation}
Here, $z^\mu (x, \xi)$ represents a point on the line connecting an appropriate 
reference point $x_0$ and the point $x$ in the 4-dimensional Minkowski space, 
with $\xi$ being the parameter that specifies the position vector $z^\mu$ 
along the path.  (To be more precise,  in the original formulation of DeWitt, 
the reference point $x_0$ is taken to be the spatial infinity. See \cite{WKZ2018}
for more detail.) It can be shown that
\begin{equation}
 \tilde{A}_\mu (x) \ = \ - \,\int_0^1 \,
 F_{\nu \sigma} (z)  \,
 \frac{\partial z^\nu}{\partial x^\mu} \,
 \frac{\partial z^\sigma}{\partial \xi} \,d \xi ,
\end{equation}
which means that $\tilde{A}_\mu (x)$ is obviously gauge-invariant,
although path-dependent.

\begin{figure}[ht]
\begin{center}
\includegraphics[width=4cm]{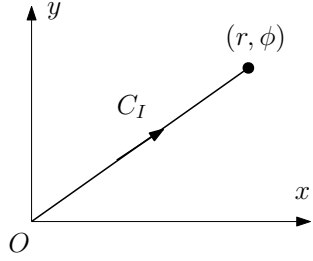}
\caption{The path $C_{\rm I}$ defined in
the circular coordinate system.}
\label{Fig:path_C_I}
\end{center}
\end{figure}

In our static problem, we can choose paths in Euclidean space with fixed time.
Let us consider time-independent axially symmetric magnetic field 
expressed as follows :
\begin{equation}
 \bm{B} (\bm{x}) \ = \ B(r) \,\bm{e}_z .
\end{equation}
A particularly simple path choice is a straight line path $C_I$ connecting
the origin and $(r, \phi)$ in the two-dimensional spherical coordinate.
(See Fig.\ref{Fig:path_C_I}.) 

\begin{figure}[ht]
\begin{center}
\includegraphics[width=11cm]{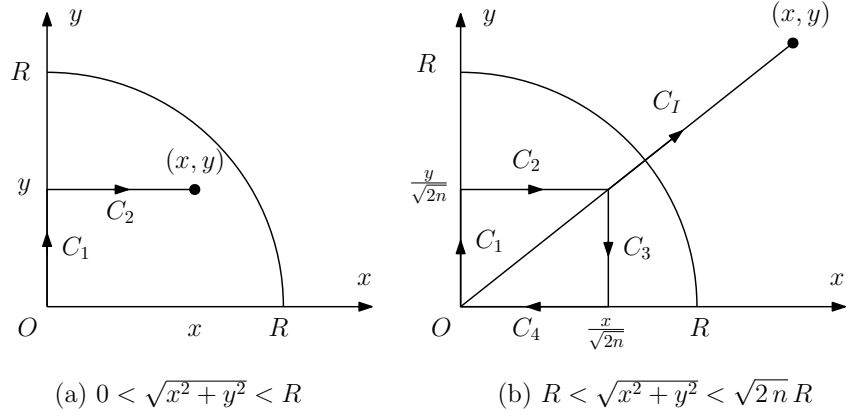}
\caption{(a) The polygonal line path $C_1 + C_2$ connecting the origin and the 
point $(x, y)$ in the rectangular coordinate system. (b) The loop path 
$C_{loop} \equiv C_1 + C_2 + C_3 + C_4$ and the straightline path $C_I$ connecting
the origin and the point $(x, y)$.}
\label{Fig:path_Landau2}
\end{center}
\end{figure}

It is easy to show that DeWitt's gauge-invariant vector potential associated 
with the path $C_I$ is given by
\begin{equation}
  \tilde{\bm{A}}^{(C_I)} (r, \phi) \ = \ 
 \left\{ \frac{1}{r} \, \int_0^r \,r^\prime \,
 B (r^\prime) \,d r^\prime \right\} \,\bm{e}_\phi ,
\end{equation}
In the particular case of AB-setting given by Eq.(\ref{B_inf_solenoid}), 
this reduces to
\begin{equation}
 \tilde{\bm{A}}^{(C_I)} (r,\phi) \ = \ \left\{
 \begin{array}{ll}
 \frac{1}{2} \,B_0 \,r \,\bm{e}_\phi \ & \ 
 (r < R) , \\
 \frac{1}{2} \,B_0 \,\frac{R^2}{r} \,\bm{e}_\phi
 \ & \ (r \geq R) , \\
 \end{array} \right.
\end{equation}
which precisely coincides with the vector potential in the symmetric
gauge. 
A natural question is what we would obtain if we choose other paths.
In particular, can we find a path which  reproduces the potential
in the generalized 2nd Landau gauge discussed before ?
The answer is simple, when the point $(x, y)$ lies inside the solenoid of 
radius $R$. In this case, we can use the knowledge already known from 
our previous study of the Landau problem \cite{WKZ2018}, and the 
answer is given by
\begin{equation}
 \tilde{\bm{A}}^{(L_2)} (x, y) \ = \ \tilde{\bm{A}}^{(C_1 + C_2)} (x, y) 
 \hspace{8mm}
 \left( \ \mbox{for} \ \ \ \sqrt{x^2 + y^2} \ < \ R \, \right) ,
\end{equation}
where $C_1 + C_2$ is a polygonal line path connecting the origin
and the point $(x, y)$ in the rectangular coordinate system as
illustrated in Fig.\ref{Fig:path_Landau2}(a). The point is that the
polygonal line path $C_1 + C_2$ is entirely contained in the domain
where the magnetic field takes a constant value $B_0$, so that 
the path-choice problem essentially reduces to the case of Landau problem 
in which the magnetic field is uniformly spread over the whole plane.

\vspace{2mm}

If the point $(x, y)$ lies outside the radius $R$ of the 
solenoid, the situation is a little complicated. First, let us assume that the point
$(x, y)$ lies outside the solenoid of radius $R$ but the point
$(x \,/\sqrt{2n}, \,y \,/\sqrt{2n})$ with $n$
being some large integer lies inside the solenoid.
In this case, we can show that the gauge potential of the generalized 2nd 
Landau gauge  is reproduced from DeWitt's gauge-invariant vector 
potential as
\begin{equation}
 \tilde{\bm{A}}^{(L_2)} (x, y) \ = \ \tilde{\bm{A}}^{(C)} (x, y) \hspace{8mm}
 \left (\ \mbox{for} \ \ \ R \ < \ \sqrt{x^2 + y^2} \ < \ \sqrt{2 \,n} \,R \,\right) ,
\end{equation}
where the path $C$ is given by
\begin{equation}
  C \ = \ C^n_{loop} \ + \ C_I .
\end{equation}
Here, $C_I$ is a straightline path directly connecting the origin and the
point $(x, y)$, while $C_{loop} \equiv C_1 + C_2 + C_3 + C_4$ is rectangular 
closed loop path as illustrated in Fig.\ref{Fig:path_Landau2}(b).
%
Finally, $C^n_{loop}$ means the $n$-fold loop integral, i.e. $n$ times 
repetition of $C_{loop}$ :
\begin{equation}
 C^n_{loop} \ \equiv \ \underbrace{C_{loop} + C_{loop} + 
 \cdots + C_{loop}}_{\tt n-times} .
\end{equation}
%


The proof goes as follows. With use of the static version of 
(\ref{A_DeWitt_1}) and (\ref{A_DeWitt_2}),
DeWitt's gauge-invariant vector potential corresponding the path $C$
and $C_I$ are respectively defined by
\begin{eqnarray}
 \tilde{\bm{A}}^{(C)} (x,y) &=& \bm{A} (x, y) \ - \ 
 \nabla \int_{C_I + C^n_{loop}} \,\bm{A} (\bm{x}^\prime) 
 \cdot d \bm{x}^\prime , \\
 \tilde{\bm{A}}^{(C_I)} (x,y) &=& \bm{A} (x, y) \ - \ 
 \nabla \int_{C_I} \,\bm{A} (\bm{x}^\prime) 
 \cdot d \bm{x}^\prime .
\end{eqnarray}
Here, $\bm{A} (\bm{x})$ is an arbitrary vector potential. Taking the
difference of the above equation, we obtain
\begin{equation}
 \tilde{\bm{A}}^{(C)} (x,y) \ - \ \tilde{\bm{A}}^{(C_I)} (x,y) \ = \ 
 - \,\nabla \,\int_{C^n_{loop}} \,\bm{A} (\bm{x}^\prime) \cdot d \bm{x}^\prime .
\end{equation}
The r.h.s. of the above equation can easily be calculated by using the
Stokes theorem : 
\begin{eqnarray}
 \int_{C^n_{loop}} \,\bm{A} (\bm{x}^\prime) \cdot d \bm{x}^\prime
 &=& n \,\int_{C_{loop}} \,\bm{A} (\bm{x}^\prime) \cdot 
 d \bm{x}^\prime \ = \ - \,n \,\int_S \,
 \left( \nabla^\prime \times \bm{A} (\bm{x}^\prime) \right) \cdot
 \bm{n} \,d S \nonumber \\
 &=& - \,n \,B_0 \,\left( \frac{x}{\sqrt{2 \,n}} \times \frac{y}{\sqrt{2 \,n}} \right)
 \ = \ - \,\frac{1}{2} \,B_0 \,x \,y . \label{loop_integral}
\end{eqnarray}
(We emphasize that the closed-loop line integral of vector 
potential is a gauge-invariant quantity.) In this way, we find that
\begin{equation}
 \tilde{\bm{A}}^{(C)} (x, y) \ = \ \tilde{\bm{A}}^{(C_I)} (x, y) \ + \ 
 \nabla \chi \hspace{5mm} \mbox{with} \hspace{5mm}
 \chi \ = \ \frac{1}{2} \,B_0 \,x \,y .
\end{equation}
The above relation holds only for $R < \sqrt{x^2 + y^2} < \sqrt{2 \,n} \,R$. 
However, since the loop-integral (\ref{loop_integral})
is independent of the integer $n$, we can extend the region of applicability
to the whole outside domain of the solenoid by taking the limit
$n \rightarrow \infty$. We are thus led to the relation
\begin{equation}
 \tilde{\bm{A}}^{(C)} (x, y) \ = \ \tilde{\bm{A}}^{(C_I)} (x, y) \ + \ 
 \bm{A}_{loop} (x, y) ,
 \hspace{8mm}
 \left( \ \mbox{for} \ \ \ R < \ \sqrt{x^2 + y^2} \, \right) ,
\end{equation}
where
\begin{equation}
 \bm{A}_{loop} (x, y) \ = \ - \,\nabla \,\lim_{n \rightarrow \infty} \,
 \int_{C^n_{loop}} \,\bm{A} (\bm{x}^\prime) \cdot d \bm{x}^\prime.
\end{equation}
This gives the relation between the vector potential in the generalized
2nd Landau gauge and that in the symmetric gauge (for 
$R < \sqrt{x^2 + y^2}$) within the DeWitt formalism. 


The argument above is just one example to show an intimate connection
between the path choice and the gauge choice. 
But, the likely existence of infinitely many
path choices seems compatible with the existence of infinitely many
gauge choices. 
Is there still any reason that favors one particular path choice ? 
As shown in our previous study of the Landau 
problem \cite{WKZ2018}, the
symmetry plays an important role in selecting a physically
favorable path in DeWitt's formulation of electrodynamics.
An advantage of the path choice $C_I$ is that it respects the
axial symmetry of the gauge potential with respect to the $z$-axis.
However, we already know that the gauge potential in the 
generalized 2nd Landau gauge, which does not
have the axial symmetry, also leads to the same AB-phase. 
Then, it appears that the symmetry requirement alone is not enough to 
uniquely fix an favorable gauge in the physics of the AB-effect, and we need some
other principle or criterion.  
It seems that the well-known Helmholtz theorem gives it.

As is widely known, the Helmholtz theorem dictates that, if the vector
potential (or any vector field) $\bm{A}$ decreases fast enough at the spatial infinity, 
it can be uniquely
decomposed into the transverse component $\bm{A}_\perp$ and
the longitudinal component $\bm{A}_\parallel$ as
\begin{equation}
 \bm{A} (\bm{x}) \ = \ \bm{A}_\perp (\bm{x})
 \ + \ \bm{A}_\parallel (\bm{x}) ,
\end{equation}          
where
\begin{eqnarray}
 \bm{A}_\perp (\bm{x}) &=& \nabla \times
 \left\{ \frac{1}{4 \,\pi} \,\int \,
 \frac{\nabla^\prime \times \bm{A} (\bm{x}^\prime)}
 {|\bm{x} - \bm{x}^\prime|} \,d^3 x^\prime \right\} , \\
 \bm{A}_\parallel (\bm{x}) &=& - \, \nabla
 \left\{ \frac{1}{4 \,\pi} \,\,\int \,
 \frac{\nabla^\prime \cdot \bm{A} (\bm{x}^\prime)}
 {|\bm{x} - \bm{x}^\prime|} \,d^3 x^\prime \right\} .
\end{eqnarray}
Each of these two components satisfies the divergenceless and 
irrotational conditions, respectively, as
\begin{equation}
 \nabla \cdot \bm{A}_\perp (\bm{x}) = 0, \hspace{10mm}
 \nabla \times \bm{A}_\parallel (\bm{x}) = 0 .
\end{equation}
Since $\nabla \times \bm{A} (\bm{x}) = \bm{B} (\bm{x})$, the
transverse component can be expressed solely with the magnetic field as
\begin{equation}
 \bm{A}_\perp (\bm{x}) = \nabla \times 
 \left\{ \frac{1}{4 \,\pi} \,\int \,
 \frac{\bm{B} (\bm{x}^\prime)}
 {|\bm{x} - \bm{x}^\prime|} \,d^3 x^\prime \right\} .
\end{equation}
This especially means that $\bm{A}_\perp (\bm{x})$ is manifestly
gauge-invariant. On the other hand, it can be shown that, under
an arbitrary gauge transformation, the longitudinal component 
$\bm{A}_\parallel (\bm{x})$ transforms in the same way as
the original vector potential $\bm{A} (\bm{x})$.
This means that $\bm{A}_\parallel (\bm{x})$ carries arbitrary
gauge degrees of freedom of the original vector potential
$\bm{A} (\bm{x})$.

Now we apply the Helmholtz theorem to our 2-dimensional problem,
where the magnetic field distribution is given by
$\bm{B} (\bm{x}) \, = \, B_0 \,\theta (R - r) \,\bm{e}_z$.
This gives
\begin{equation}
 \bm{A}_\perp (\bm{x}) \ = \ - \,\nabla \times \left\{
 \frac{B_0 \,\bm{e}_z}{2 \,\pi} \,\iint \,
 d x^\prime \,d y^\prime \,\theta (R - r^\prime) \,
 \ln \left( |\bm{x} - \bm{x}^\prime | \,/\, r_0 \right) 
 \right\} ,
\end{equation}
where $r_0$ is some constant with the dimension of length.
After carrying out the integral, we find that \cite{Adachi1992}
\begin{eqnarray}
 {\bm{A}}_\perp (\bm{x}) \ = \ \left\{
 \begin{array}{ll}
 \frac{1}{2} \,B_0 \,r \,\bm{e}_\phi \ & \ 
 (r < R) , \\
 \frac{1}{2} \,B_0 \,\frac{R^2}{r} \,\bm{e}_\phi
 \ & \ (r \geq R) , \\
 \end{array} \right.
\end{eqnarray}
which is nothing but the vector potential in the symmetric gauge.
Thus, we observe that, under the condition of the Helmholtz theorem to hold, 
the transverse part of the vector potential corresponding to the infinitely 
long solenoid is uniquely fixed and it just coincides with
the vector potential in the symmetric gauge.
Since the transverse part of the vector potential is gauge-invariant, 
some authors propose to regard it as 
``physical component'' \cite{Adachi1992},\cite{Stewart2003},\cite{Li2012}.
According to these authors, the AB-effect is due to the transverse component 
of the vector potential, which should be regarded as a
``physical reality''.
In fact, the longitudinal part $\bm{A}_\parallel (\bm{x})$ never contributes
to the AB-phase : 
\begin{eqnarray}
 \phi_{AB} \ = \ e \,\oint \,\left(
 \bm{A}_\perp (\bm{x}) + \bm{A}_\parallel (\bm{x})
 \right) \cdot d \bm{x} \ = \ 
 e \,\oint \,\bm{A}_\perp (\bm{x}) \cdot d \bm{x}
 \ = \ e \, \Phi ,
\end{eqnarray}
since
\begin{eqnarray}
 \oint \,\bm{A}_\parallel (\bm{x}) \cdot d \bm{x}
 \ = \ \int_S \, \nabla \times 
 \bm{A}_\parallel (\bm{x}) 
 \cdot \bm{n} \, d S \ = \ 0 ,
\end{eqnarray}
due to the defining property of the longitudinal component
$\nabla \times \bm{A}_\parallel = 0$.
Note that, in this flow of logic, the vector potential in the outer region is determined 
by the magnetic field at a remote place, i.e. the field
inside the solenoid.
Does it then support the non-local interpretation of the AB-effect ?
We shall come back to this question shortly.

At this point, we
believe it useful to recognize an intimate connection between the
Helmholtz decomposition and the Coulomb gauge fixing, although
they are not completely equivalent. 
Suppose that we take the Coulomb gauge vector potential
satisfying $\nabla \cdot \bm{A} = 0$. 
Putting it into the Helmholtz decomposition, we obtain
\begin{eqnarray}
 \bm{A} (\bm{x}) \ = \ 
 \bm{A}_\perp (\bm{x}) 
 \ = \ 
 \nabla \times \frac{1}{4 \,\pi} \,\int \,
 \frac{\bm{B} (\bm{x}^\prime)}{|\bm{x} - \bm{x}^\prime|} \,
 d^3 x^\prime , \hspace{12mm}
 \bm{A}_\parallel (\bm{x}) = 0 .
\end{eqnarray}
The elimination of the longitudinal component means that the gauge
is completely fixed. In the stationary problem, one of the Maxwell equations
reduces to
\begin{equation}
 \nabla \times \bm{B} = \mu_0 \,\bm{J},
\end{equation}
where $\bm{J}$ is the electric current and $\mu_0$ is the magnetic
permeability of the vacuum. Then, the transverse part of the vector
potential can also be expressed as
\begin{eqnarray}
 \bm{A} (\bm{x}) \ = \ \frac{\mu_0}{4 \,\pi} \,
 \nabla \times \int \,\frac{\bm{J} (\bm{x}^\prime)}
 {|\bm{x} - \bm{x}^\prime|} \,d^3 x^\prime .
\end{eqnarray}
This means that the vector potential is expressed solely with the electric current, 
which is undoubtedly a ``physical reality''.

The observed intimate connection between the Helmholtz decomposition
and the Coulomb gauge raises another question.
Is the transverse component of the vector potential obtained from the Helmholtz 
theorem really a gauge-invariant quantity in more general or wider sense ?
Here we are thinking of more general time-dependent gauges
like the Lorentz gauge.
To answer this question, it is useful to remember the generalization of the 
Helmholtz theorem in the 4-dimensional Minkowski 
space-time \cite{Kobe1984},\cite{Heras2016}.
It is known that the 4-dimensional generalization of the Helmholtz is not 
necessarily unique, but practically most useful one is the following.
It is called the causal Helmholtz theorem for time dependent vector
potential $\bm{A} (\bm{x}, t)$ and given in the form \cite{Heras2016} :
\begin{eqnarray}
 \bm{A} (\bm{x}, t) \! \! &=& - \,\nabla \,
 \int \,\frac{[\nabla^\prime \cdot \bm{A}]}{4 \,\pi R} \,
 d^3 x^\prime \ + \ \nabla \times 
 \int \,\frac{[\nabla^\prime \times \bm{A}]}{4 \,\pi R} \,
 d^3 x^\prime \ + \ \frac{1}{c^2} \,\frac{\partial}{\partial t} \,
 \int \,\frac{\left[ \frac{\partial}{\partial t} \,\bm{A} \right]}
 {4 \,\pi R} \,d^3 x^\prime , \ \  \label{causal_Helmholtz_theorem}
\end{eqnarray}
where $R = |\bm{x} - \bm{x}^\prime|$, while the brace notation $[\, ]$
means that the enclosed quantity is to be evaluated at the retarded time                         
$t^\prime = t - R / c$ with $c$ being the light velocity.
(In the argument of the causal Helmholtz theorem, we use the SI unit
instead of the natural unit.) 
The causal Helmholtz theorem above means that, the time-dependent
vector potential is decomposed into three pieces, each of which is
specified by its divergence, rotation, and time derivative, respectively.
Just like that the standard Helmholtz theorem is simplified by taking
the Coulomb gauge, the causal Helmholtz theorem is simplified if
we take the Lorentz gauge specified by the condition \cite{Heras2016} : 
\begin{eqnarray}
 \nabla \cdot \bm{A} \ + \ 
 \frac{1}{c^2} \frac{\partial}{\partial t} \,\phi
 \ = \ 0 .
\end{eqnarray} 
First, by noting that
\begin{eqnarray}
 - \,\nabla \,\int \,\frac{[\nabla^\prime \cdot 
 \bm{A}]}{4 \,\pi R} \,d^3 x^\prime \ = \ 
 \nabla \,\int \,\frac{\left[ \frac{1}{c^2} \,
 \frac{\partial}{\partial t} \,\phi \right]}{4 \,\pi \,R} \,
 d^3 x^\prime \ = \ 
 \frac{1}{c^2} \,\frac{\partial}{\partial t} \,
 \int \,\frac{[\nabla^\prime \phi]}{4 \,\pi \,R} \,d^3 x^\prime ,
\end{eqnarray}
we get
\begin{eqnarray}
 \bm{A} \ = \ \frac{1}{4 \,\pi \,c^2} \,
 \frac{\partial}{\partial t} \,\int \,
 \frac{\left[ \nabla^\prime \phi + 
 \frac{\partial}{\partial t} \,\bm{A} \right]}{R} \,d^3 x^\prime
 \ + \ \nabla \times \int \,\frac{[\bm{B}]}{4 \,\pi \,R} \,d^3 x^\prime .
\end{eqnarray}
Next, using one of the Maxwell equations
$\nabla \times \bm{B} = \mu_0 \,\bm{J} + \epsilon_0 \,\mu_0 \,
\frac{\partial}{\partial t} \,\bm{E}$, where $\epsilon_0$ and
$\mu_0$ are respectively the electric permittivity and the
magnetic permeability of the vacuum, we have
\begin{eqnarray}
 \bm{A} \ = \ \frac{1}{4 \,\pi c^2} \,
 \frac{\partial}{\partial t} \,\int \,
 \frac{\left[ \nabla^\prime \phi + 
 \frac{\partial}{\partial t} \,\bm{A} + 
 \bm{E} \right]}{R} \,d^3 x^\prime \ + \ 
 \frac{\mu_0}{4 \,\pi} \,\int \,
 \frac{[\bm{J}]}{R} \,d^3 x^\prime .
\end{eqnarray}
Here, we have used the relation $\epsilon_0 \,\mu_0 = 1 / c^2$.
Finally, using the relation $\bm{E} = - \,\nabla \phi - 
\frac{\partial}{\partial t} \, \bm{A}$, we obtain
\begin{eqnarray}
 \bm{A} (\bm{x}, t) \ = \ \frac{\mu_0}{4 \,\pi}  \,
 \int \,\frac{[\bm{J}]}{R} \,d^3 x^\prime \ = \ 
 \frac{\mu_0}{4 \,\pi} \,\int \,
 \frac{\bm{J} \left(\bm{x}^\prime, t - 
 \frac{|\bm{x} - \bm{x}^\prime|}{c}\right)}
 {|\bm{x} - \bm{x}^\prime|} \,d^3 x^\prime .
\end{eqnarray}
This is nothing but the familiar retarded potential obtained in the
Lorentz gauge.

Coming back to the causal Helmholtz theorem (\ref{causal_Helmholtz_theorem}), 
we note that the magnetic field in the AB-effect measurement is static, 
i.e. $\bm{A} (\bm{x}, t) \ = \ \bm{A} (\bm{x})$, so that
\begin{eqnarray}
 [\nabla^\prime \cdot \bm{A}] \ = \ \nabla^\prime \cdot \bm{A}
 (\bm{x}^\prime), \ \ 
 [\nabla^\prime \times \bm{A}] \ = \ \nabla^\prime \times \bm{A}
 (\bm{x}^\prime), \ \ 
 \left[ \frac{\partial}{\partial t} \bm{A} \right] \ = \ 0 .
\end{eqnarray}
The causal Helmholtz theorem then reduces to the standard Helmholtz theorem. 
Since, in the argument above, the standard Helmholtz theorem is obtained
as a limiting case of the causal Helmholtz theorem, which meets the causality 
requirement, the seemingly non-locality feature of the Helmholtz decomposition 
should not be taken as problematical.
This means that the reality interpretation of the vector potential and
the non-locality interpretation of the AB-effect is not necessarily
a contradictory idea.

Still puzzling in the AB-effect is the physical meaning as well as
the role of non-zero vector potential in the completely magnetic
field free region.
As we shall see, this oddness of the pure-gauge potential
manifests most drastically in the
orbital angular momentum (OAM) of the electron moving outside
the solenoid.

\section{The role of electron orbital angular momentum in the
AB-effect}
\label{Section:s3}

Our discussion here starts with the vector potential corresponding to
the confined uniform magnetic field inside an infinitely-long solenoid
with radius $R$ in the symmetric gauge, which
can be identified with the transverse component of the gauge potential
in arbitrary gauge,
\begin{eqnarray}
 \bm{A}_\perp (\bm{x}) \ = \ \left\{
 \begin{array}{ll}
 \frac{1}{2} \,B_0 \,r \,\bm{e}_\phi \ \ & \  
 (r < R) , \\
 \frac{1}{2} \,B_0 \,\frac{R^2}{r} \,\bm{e}_\phi
 \ \ & \ 
 (r \geq R) . \\
 \end{array} \right.
\end{eqnarray}
We recall that the vector potential outside the solenoid takes
the so-called pure-gauge form : 
\begin{eqnarray}
 \bm{A}^{out}_\perp (\bm{x}) \ = \ \nabla \,
 \left( \frac{1}{2} \,B_0 \,R^2 \,\phi \right) ,
\end{eqnarray}
as naturally anticipated from the magnetic-field-free condition 
$0 = \bm{B}^{out} = \nabla \times \bm{A}^{out}_\perp$. Curiously, as was
emphasized by Tiwari in a recent paper \cite{Tiwari2017}, 
this pure-gauge potential appears to carry non-zero orbital 
angular momentum : 
\begin{eqnarray}
 L^{out}_z &=& e \,\left( \bm{x} \times 
 \bm{A}^{out}_\perp \right)_z 
 \ = \ \frac{1}{2} \,e \,B_0 \,R^2 \ = \  
 \frac{e}{2 \,\pi} \,\Phi .
\end{eqnarray}

We first point out that this OAM just corresponds to the 
``potential angular momentum'' in the terminology of the
papers \cite{Wakamatsu2010},\cite{Wakamatsu2011}, which
discuss the gauge-invariant decomposition of the nucleon spin
as well as the gauge-invariant decomposition of the
photon angular momentum. (The reviews of the nucleon spin
decomposition problem can, for example, be found in
\cite{Leader-Lorce2014}, \cite{Wakamatsu2014}.)
In general, our potential angular momentum is defined by
\begin{eqnarray}
 L^{pot}_z \ = \ e \,(\bm{x} \times 
 \bm{A}_\perp)_z , \label{Lpot_def}
\end{eqnarray}
where $\bm{A}_\perp$ is the transverse component of the
vector potential.
Since $\bm{A}_\perp$ is a gauge-invariant quantity, the potential
angular momentum defined as above is also a gauge-invariant quantity.
To understand the physical meaning of the potential angular momentum,
it is instructive to remember the origin of this 
quantity within the framework of general electromagnetic theory of
angular momentum \cite{Wakamatsu2010},\cite{Cohen-Tannoudji1989}.
As is widely-known, the total angular momentum of the photon or
the electromagnetic field is given by
\begin{equation}
 \bm{J}^\gamma \ = \ \epsilon_0 \,
 \int \,\bm{x}^\prime \times \left( \bm{E} (\bm{x}^\prime)
 \times \bm{B} (\bm{x}^\prime) \right) \,d^3 x^\prime . \label{J_gamma}
\end{equation}
By using the transverse-longitudinal decomposition of the vector
potential $\bm{A} = \bm{A}_\perp + \bm{A}_\parallel$,
the electric field can also be decomposed into the transverse and
longitudinal components as
\begin{equation}
 \bm{E} \ = \ \bm{E}_\perp \ + \ \bm{E}_\parallel , 
\end{equation}
where
\begin{equation}
 \bm{E}_\perp (\bm{x}) \ = \ - \,\frac{\partial \bm{A}_\perp}{\partial t},
 \hspace{10mm}
 \bm{E}_\parallel \ = \ - \,\nabla \phi \ - \ 
 \frac{\partial \bm{A}_\parallel}{\partial t} .
\end{equation}
On the other hand, the magnetic field has only the transverse component,
i.e. $\bm{B} = \bm{B}_\perp$, 
since $\nabla \times \bm{A}_\parallel = 0$ by definition.
The total angular momentum of the electromagnetic field can therefore be
decomposed into two pieces as
\begin{equation}
 \bm{J}^\gamma \ = \ \epsilon_0 \,
 \int \,\bm{x}^\prime \times \left( \bm{E}_\perp (\bm{x}^\prime)
 \times \bm{B} (\bm{x}^\prime) \right) \,d^3 x \ + \ \epsilon_0 \, 
 \int \,\bm{x}^\prime \times \left( \bm{E}_\parallel (\bm{x}^\prime)
 \times \bm{B} (\bm{x}^\prime) \right) \,d^3 x^\prime . \label{Jgamma_TL}
\end{equation}
The potential angular momentum originates from the term
containing $\bm{E}_\parallel$ in Eq.(\ref{Jgamma_TL}), i.e.
\begin{eqnarray}
 \epsilon_0 \,\int \bm{x}^\prime \times \left(
 \bm{E}_\parallel (\bm{x}^\prime) \times 
 \bm{B} (\bm{x}^\prime) \right) \,d^3 x^\prime
 &=& \int \,\rho (\bm{x}^\prime) \,(\bm{x}^\prime \times 
 \bm{A}_\perp (\bm{x}^\prime) ) \,\,d^3 x^\prime \ + \ \mbox{S.T.} \nonumber \\
 &=& \hspace{8mm} e \,(\bm{x} \times \bm{A}_\perp (\bm{x}))  \ + \ 
 \mbox{S.T.} \label{L_pot}
\end{eqnarray}
Here, S.T. stand for surface integral terms.
For obtaining the above equation, we have used the Gauss law 
$\nabla \cdot \bm{E}_\parallel = \rho / \epsilon_0$ as well as the
relation $\rho (\bm{x}^\prime) = e \,\delta (\bm{x}^\prime - \bm{x})$
with $\bm{x}$ being the position vector of the electron.
The quantity $\bm{L}^{pot} (\bm{x}) \equiv e \,(\bm{x} \times \bm{A}_\perp (\bm{x}))$
appearing in the r.h.s. is just our potential angular momentum, which
can be interpreted as the angular momentum associated with the
longitudinal component of the electromagnetic field 
generated by charged particle sources.
(We recall that the terminology potential angular momentum is a
straightforward generalization of the potential momentum
$e \bm{A} (\bm{x})$ advocated by Konopinski \cite{Konopinski1978},\cite{Wakamatsu2010}.) 
An important observation is that, since  $\bm{E}_\parallel$
satisfies the Gauss law $\nabla \cdot \bm{E}_\parallel = \rho / \epsilon_0$,
the potential angular momentum vanishes in the absence of
charged particle sources.
A delicate question is then which of charged particles or photons should it be
attributed to. It is of the same sort of question as which of 
charged particles or electromagnetic fields should the Coulomb energy
be attributed to. To attribute it to charged particles is closer to the
concept of ``action at a distance principle'', while to attribute it to
an electromagnetic fields is closer to the concept of ``action
through medium principle''. If there is no difference between their
physical predictions, the choice is just a matter of convenience.

Although it has little to do with our discussion below, 
one might be interested also in the physical meaning of the first
term of Eq.(\ref{Jgamma_TL}), which contains the transverse part
of the electric field $\bm{E}_\perp$. It is known that
this first term can further be decomposed into two pieces 
as \cite{Wakamatsu2010},\cite{Cohen-Tannoudji1989}
\begin{eqnarray}
 \epsilon_0 \,\int \bm{x}^\prime \times \left(
 \bm{E}_\perp (\bm{x}^\prime) \times \bm{B} (\bm{x}^\prime) \right) \,d^3 x^\prime
 &=& \epsilon_0 \,
 \int \bm{E}_\perp (\bm{x}^\prime) \times \bm{A}_\perp (\bm{x}^\prime)
 \,d^3 x^\prime \nonumber \\
 &+& \epsilon_0 \,
 \int E^j_\perp (\bm{x}^\prime) \,(\bm{x}^\prime \times \nabla^\prime) \,
 A^j_\perp (\bm{x}^\prime) \,d^3 x^\prime  \ + \ \mbox{S.T.}, \label{SO_photon}
\end{eqnarray}
up to the surface integral terms abbreviated as S.T.
The two pieces on the r.h.s. respectively correspond to the intrinsic
spin and orbital angular momentum of a real photon.
An important difference from the potential angular momentum
associated with $\bm{E}_\parallel$ is that the two terms in (\ref{SO_photon})
survive even if the interaction between charged particles and 
electromagnetic fields is turned off.

In the most commonly encountered situation, in which the motion of
charged particles are confined in a localized domain in the 3-dimensional space, 
the surface terms in (\ref{L_pot}) and (\ref{SO_photon}) are known to vanish 
simply and they do not play any remarkable role.
However, this is not the case for our Aharonov-Bohm solenoid problem
with 2-dimensional geometry. It can easily be convinced from the
2-dimensional version of (\ref{L_pot}). 
Let us consider the spatial region outside the solenoid.
In this region, there is no magnetic field, so that
the l.h.s. of (\ref{L_pot}) is expected to vanish identically. 
On the other hand, the potential angular momentum
in the r.h.s. is obviously nonzero, as we have pointed out at the beginning.
Undoubtedly, this puzzling observation is intimately connected with
our raised question of why the pure-gauge configuration appears to
bear non-zero orbital angular momentum.  
From the mathematical point of view, only way out of this dilemma 
would be that the surface term on the r.h.s. of (\ref{L_pot}) does not vanish.
Later, we shall verify that this is in fact so. For the time being, we shall 
continue the discussion of more physical nature.

An important general relation to remember is 
that \cite{Wakamatsu2010},\cite{Wakamatsu2011}
\begin{eqnarray}
 L^{g.i.c.}_z \ = \ L^{mech}_z \ + \ L^{pot}_z ,
\end{eqnarray}
where  $L^{mech}_z$ is the standard ``mechanical OAM'' of the electron
defined by
\begin{eqnarray}
 L^{mech}_z \ = \ e \,(\bm{x} \times 
 \bm{\Pi} )_z \ = \ 
 e \,\left[ \bm{x} \times (\bm{p} - e \,
 \bm{A} ) \right]_z ,
\end{eqnarray}
while $L^{g.i.c.}_z$ is the so-called ``gauge-invariant-canonical OAM'',
the concept of which was introduced by 
Chen et al. \cite{Chen2008},\cite{Chen2009},
\begin{eqnarray}
 L^{g.i.c.}_z \ = \ 
 e \,\left[ \bm{x} \times (\bm{p} - e \,
 \bm{A}_\parallel ) \right]_z .
\end{eqnarray}

Now, remembering that
\begin{eqnarray}
 \bm{A}_\perp (\bm{x}) &=& \frac{1}{2} \,B_0 \,
 \frac{R^2}{r} \,\bm{e}_\phi \ = \ A_\phi (r) \,
 \bm{e}_\phi , \\
 L^{pot}_z &\equiv& e \,(\bm{x} \times \bm{A}_\perp)_z
 \ = \ e \,r \,A_\phi  (r),
\end{eqnarray}
an inseparable connection between the AB-phase and the potential 
angular momentum becomes clear from the relation
\begin{eqnarray}
 \phi_{AB} 
 &=& e \,\oint \,\bm{A}_\perp (\bm{x}) \cdot
 d \bm{x} \ = \ e \,\int_0^{2 \,\pi} \,A_\phi (r) \,r \,d \phi 
 \ = \ 2 \,\pi \, L^{pot}_z .
\end{eqnarray}
To our knowledge, this relation has never been written down before
at least in the present form.
Since the potential angular momentum is a gauge-invariant quantity, 
the above  is a gauge-invariant relation, which holds independently of the gauge choice. 
This owes to gauge-invariant nature of our potential angular momentum,
which contains $\bm{A}_\perp$ instead of $\bm{A}$.

A natural question is what is the physical or dynamical origin of non-zero
OAM in the region where the vector potential is of pure-gauge form.
In a recent paper \cite{Tiwari2017}, Tiwari advocated answering
this question based on the idea of modular angular momentum exchange.
The idea of {\it modular variable} in quantum physics was first 
introduced by Aharonov, Pendleton and Petersen \cite{APP1969}, 
with a special intention of showing non-local nature of the
AB-effect as well as its purely quantum mechanical nature without
any classical analog. Formally, the modular variable is defined as a 
continuous physical quantity modulo its basic unit.
For example, the modular momentum $p_x \,(\mbox{mod} \ p_0)$ is defined as
\begin{equation}
 p_x \,(\mbox{mod} \ p_0) = \ p_x \ - \ N \,p_0 .
\end{equation}
Here $N$ is an integer, and $p_0 = h \,/\,L$ is a basis unit defined by the 
Plank constant $h$ and some constant spatial length $L$. 
They claim that the potential effect in the AB-effect can be viewed as
exchange of a non-local dynamical quantity, the modular momentum,
with no exchange of local quantities.
Later, Aharonov and Kaufherr pushed forward this idea to carry
out more concrete analysis of the Aharonov-Bohm solenoid
effect \cite{AK2004}. 
According to these authors, the Aharonov-Bohm solenoid
effect can be explained as either of the exchange of modular momentum
or modular angular momentum between the electron and
the solenoid. Moreover, their theoretical
analysis provided a nontrivial answer to the interesting question
about when and where this modular variable exchange occurs.
Unfortunately, the answer appears to depend on the choice of gauge, 
although the final answer to the AB-phase shift is gauge-independent.
This might not necessarily be a serious problem, since the concept of
modular variable is of purely quantum mechanical nature, and
there is no clear idea of particle trajectory in quantum
mechanics.

Still, it would be nice if we can answer the proposed question above by 
using the standard orbital angular momentum concept not the sophisticated 
concept of modular angular momentum. The reason is partly because it has an 
advantage of keeping contact with the classical analog. 
Curiously, one can provide two totally different explanations based on 
two totally different mechanisms or local forces.  

\begin{itemize}
\item Explanation based on the induced electric field of time-varying
 magnetic flux

\item Explanation based on the magnetic Lorentz force due to a
finite-length solenoid, and the subsequent operation of
infinite-length limit.

\end{itemize}

As we shall see, the existence of two totally different or complimentary 
explanations is related to a singular nature of the idealized setting of the 
AB-phase shift measurement, which comes from the assumption of infinitely
long solenoid.

\subsection{Explanation by the induced electric field}

A novel observation by Peshkin \cite{Peshkin1981} and also by
Miyazawa-Miyazawa \cite{Miyazawa} is the following.
Although the magnetic field inside the solenoid cannot be seen from the outside, 
if the magnetic field inside the solenoid varies time-dependently, induced
electric field is generated even in the outside region, so that the electron
can feel it.
Let us assume that the magnetic field inside an infinitely long solenoid
is uniform but time-dependent, and it is given by
\begin{eqnarray}
 \bm{B} (\bm{x}, t) &=& B(t) 
 \,\theta (R-r) \,\bm{e}_z .
\end{eqnarray}
The corresponding transverse part of the vector potential outside
the solenoid is given by
\begin{equation}
 \bm{A}^{out}_\perp (\bm{x}, t) \ = \ \frac{1}{2} \,
 B(t)  \,
 \frac{R^2}{r} \,\bm{e}_\phi \,.
\end{equation}
Once the time-dependence of the magnetic field is introduced, one
might be worried about the retardation effect. The retardation effect
of the vector potential induced by the surface current of the solenoid,
which is turned on suddenly at some time, was discussed in some detail 
in the paper by Peshkin \cite{PT1989}. 
To avoid inessential complexity, we use here the simplified treatment, 
in which the retardation effect does not play any crucial role.
The simplest justification of this approach would be obtained by assuming 
adiabatically slow variation of the magnetic field inside the solenoid. 
This then enables us to treat the problem basically as a quasi-static 
one.

The induced electric field generated by the above time-dependent vector
potential is given by
\begin{eqnarray}
 \bm{E} \ = \ \bm{E}^{out} \ = \ - \,
 \frac{\partial \bm{A}^{out}_\perp}{\partial t} \ = \ 
 - \,\frac{1}{2} \,\dot{B} (t) 
 \,\frac{R^2}{r} \,
 \bm{e}_\phi \ \equiv \ E_\phi (r) \,\bm{e}_\phi .
\end{eqnarray}
Note that this induced electric field is oriented to the
circumferential direction.

Consider an electron located at the distance $r \,(\,> R)$ from 
the solenoid center. This electron feels torque
\begin{eqnarray}
 N \ = \ e \,r \,E_\phi (r) .
\end{eqnarray}
The equation of motion for the electron is therefore given by
\begin{eqnarray}
 \frac{d}{d t} \, L^{mech}_z 
 \ = \ e \,r \,E_\phi (r) .
\end{eqnarray}
We emphasize that what appears in equation of motion is the mechanical OAM,  
not the canonical OAM.
Making use of the axial symmetry of the induced electric field, we get
\begin{eqnarray}
 \frac{d}{d t} \,L^{mech}_z &=& e \,r \,E_\phi (r)
 \ = \ \frac{e}{2 \,\pi} \,\oint_{|\bm{x}| = r} \,
 \bm{E} (\bm{x}) \cdot  d \bm{x} \nonumber \\
 &=& \ - \,\frac{e}{2 \,\pi} \,\oint_{|\bm{x}| = r} \,
 \frac{\partial}{\partial t} \,\bm{A} \cdot d \bm{x}
 \ = \ - \,\frac{e}{2 \,\pi} \,\frac{d}{d t} \,
 \oint_{|\bm{x}| = r} \,\bm{A} \cdot d \bm{x} \nonumber \\
 &=& - \,\frac{e}{2 \,\pi} \,\frac{d}{d t} \,
 \iint_{|\bm{x}| < r} \,(\nabla \times \bm{A}) \cdot
 \bm{n} \,d S \ = \ 
 - \,\frac{e}{2 \,\pi} \,\frac{d}{d t} \,\Phi (t) .
\end{eqnarray}
Integrating out the obtained equation with the initial and final
conditions : 
\begin{eqnarray}
 B(t) = \left\{ \begin{array}{ll}
 0 \ & \ \mbox{\tt when} \ t \leq 0 ,\\
 B_0 \ & \ \mbox{\tt when} \ t \geq t_f ,\\
 \end{array} \right.
\end{eqnarray}
we obtain
\begin{eqnarray}
 L^{mech}_z (t_f) \ - \ L^{mech}_z (0) \ = \ 
 - \,\frac{e}{2 \,\pi} \,\left[ \Phi (t_f) \ - \ 
 \Phi (0) \right] .
\end{eqnarray}
Since $\Phi (0) = 0$, and $\Phi (t_f) = \pi \,R^2 \,B_0 = \Phi$, 
we eventually arrive at the relation
\begin{eqnarray}
 L^{mech}_z (t_f) \ = \ L^{mech}_z (0) \ - \ \beta ,
\end{eqnarray}
with $\beta = e \,\Phi / 2 \,\pi$.
The relation obtained above is completely consistent with our general
relation $L_z^{g.i.c.} = L_z^{mech} + L_z^{pot}$.
It can be seen as follows. First, since the magnetic flux at $t=0$
is zero, $L_z^{mech} (0)$ can be identified with the canonical OAM,
which is denoted as $L_z^{g.i.c.}$ in our gauge-invariant formulation.  
Second. $\beta = e \,\Phi / 2 \,\pi$ just coincides with the
potential OAM $L_z^{pot}$. This means that the relation
$L_z^{g.i.c.} = L_z^{mech} + L_z^{pot}$ holds.
The interpretation of this relation is therefore as follows.
When the magnetic flux inside the solenoid is increased from 0
to $\Phi$, the mechanical OAM changes from $L_z^{g.i.c.}$
to $L_z^{g.i.c.} - \beta$, whereas the potential OAM changes
from 0 to $\beta$. As a consequence, the canonical OAM
remains a constant throughout this process. 

In any case, we emphasize the fact that the above change of the electron
mechanical OAM is caused by the induced electric field, which acts
locally (or directly) on the electron outside the solenoid.
Remembering that the AB-phase shift is related to the potential OAM as
\begin{eqnarray}
 \phi_{AB} \ = \ e \,\oint \,\bm{A}_\perp \cdot d \bm{x}
 \ = \ 2 \,\pi \,L^{pot}_z ,
\end{eqnarray}
together with the relation
\begin{eqnarray}
 \Delta L^{mech}_z \ \equiv \ L^{mech}_z (t_f) \ - \ 
 L^{mech}_z (0) \ = \ - \, L^{pot}_z  ,
\end{eqnarray}
one may be able to say that the cause of AB-phase shift can be 
associated with the change of the electron mechanical OAM generated
by the induced electric field, which acts {\it locally} on the electron
in the magnetic-field free region. Unfortunately,
although the above explanation by the time-varying magnetic field grasps some 
essence of the AB-effect, it does not faithfully correspond to the
actual setup of the AB-effect measurement.
This is because, in the actual setting of the AB-phase measurements, 
the electron beam is projected only after the magnetic field is turned on and
kept constant.

\subsection{Explanation by the magnetic Lorentz force}

First, let us consider a solenoid with finite length $2 L$ and radius $R$.
The behavior of the magnetic field around the finite length solenoid
was investigated by Babiker and Lowdon \cite{BL1984}.
As shown there, the magnetic field is positive inside the solenoid,
while it is negative well outside the solenoid. 
The latter property is of course due to the return magnetic flux from
the finite-length solenoid.
As shown in their paper, at very large distance, the vector potential
and the magnetic field of the finite-length solenoid takes the
following approximate forms : 
\begin{eqnarray}
 \left\{ \begin{array}{l}
 A_\phi (r) \ \simeq \ \ \ \frac{1}{2} \,B_0 \,
 \frac{R^2 \,L}{r^2} \\
 B_z (r) \ \simeq \ -  \,
 \frac{1}{2} \,B_0 \,\frac{R^2 \,L}{r^3} \\
 \end{array} \right.
 \hspace{16mm} ( r \gg R, \,\,r \gg L) .
\end{eqnarray}
Note that the above vector potential of the finite-length solenoid
damps as $1 / r^2$, in contrast to the fact the vector potential
of the infinite-length solenoid decreases as $1 / r$.
Then, the circular integral of the vector potential around a circle with
very large radius vanishes as $1/r$, which means that
\begin{eqnarray}
 \lim_{r \rightarrow \infty} \,2 \,\pi \,r \,A_\phi (r)
 \ = \ \lim_{r \rightarrow \infty} \,
 \int_{|\bm{x}| = r} \,\bm{A} (\bm{x}) \cdot d \bm{x}
 \ = \ 0 .
\end{eqnarray}
The Stokes theorem then dictates that the total magnetic flux through
$z = 0$ plane vanishes. The fact is that the return flux exactly cancels
the magnetic flux inside the solenoid.
This is just consistent with the solenoidal feature of the magnetic flux, 
which obeys the rule $\nabla \cdot \bm{B} = 0$.
In fact, the following identity must always hold for
the magnetic field, which is generated by the electric current
according to the Biot-Savart law,
\begin{eqnarray}
 \oint \,\bm{A} (\bm{x}) \cdot d \bm{x} &=& 
 \int_S (\nabla \times \bm{A} (\bm{x}) ) \cdot \bm{n} \,d S \nonumber \\
 &=& \int_S \,\bm{B} (\bm{x}) \cdot \bm{n} \,d S \ = \ 
 \int_V \,\nabla \cdot \bm{B} (\bm{x}) 
 \, d V \ = \ 0 .
\end{eqnarray}
It should be contrasted with the case, in which  large $L$ limit is taken first : 
\begin{eqnarray}
 \left\{ \begin{array}{l}
 A_\phi (r) \ \simeq \ \frac{1}{2} \,B_0 \,
 \frac{R^2}{r} \ = \ 
 \frac{\Phi}{2 \,\pi \,r} \\
 B_z (r) \ \simeq \ \ \ B_0 \,
 \frac{R^3}{r^3} \\
 \end{array} \right.
 \hspace{10mm} ( L \rightarrow \infty, \ r \gg R) .
\end{eqnarray}
In this case, one has
\begin{eqnarray}
 \lim_{r \rightarrow \infty} \,2 \,\pi \,r \,A_\phi (r)
 \ = \ \lim_{r \rightarrow \infty} \,
 \int_{|\bm{x}| = r} \,\bm{A} (\bm{x}) \cdot d \bm{x}
 \ = \ \Phi .
\end{eqnarray}
This simple comparison already indicates quite singular nature of 
infinitely long solenoid as an idealized object, for which there is
no return magnetic flux.
In the following, we investigate the behavior of electron OAM 
around the solenoid with finite length $2 L$.
Then, we shall see what happens in the $L \rightarrow \infty$ limit.

\begin{figure}[ht]
\begin{center}
\includegraphics[width=6cm]{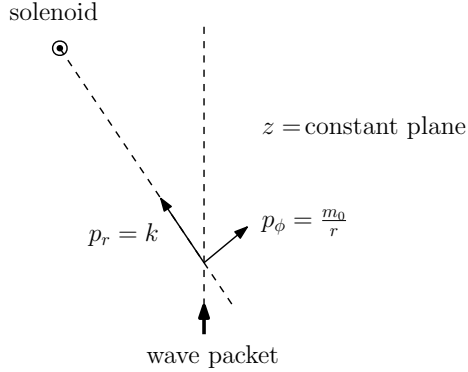}
\caption{A cylindrical wave packet is approaching the solenoid
in a $z = $ constant plane with a certain momentum, the asymptotic
value of which is specified
by negative radial component $p_r = k$ and a certain angular
component $p_\phi = \frac{m_0}{r}$.}
\label{Fig:finite_solenoid}
\end{center}
\end{figure}

Following Olariu and Popescu \cite{GP1985}, let us assume that
a cylindrical wave packet is approaching the solenoid in a
$z = $ constant plane with a certain momentum, the asymptotic
(large $r$) value of which is specified by negative radial
component $p_r = k$ and angular component
$p_\phi = \frac{m_0}{r}$ with $m_0$ being some real constant.
(See Fig.\ref{Fig:finite_solenoid}.)

\vspace{5mm}

Crucial point here is the existence of non-zero {\it return magnetic flux}
in the outside region of the finite-length solenoid, which exerts 
Lorentz force on the electron moving outside the solenoid.
The equation of motion for the electron mechanical OAM
$\bm{L}^{mech} \ = \ L^{mech}_z (r,z) \,\bm{e}_z$, is given by
\begin{eqnarray}
 \frac{d}{d t} \,\bm{L}^{mech} &=& \bm{N} \ = \ 
 \bm{x} \times e \,
 (\bm{v} \times \bm{B}) \nonumber \\
 &=& e \,r \,\bm{e}_r \times \left(
 (v_r \,\bm{e}_r \ + \ v_\phi \,\bm{e}_\phi) 
 \times B_z (r, z) \,\bm{e}_z \right) \nonumber \\
 &=& - \,e \,r \,v_r \,\,B_z (r,z) \,\bm{e}_z .
\end{eqnarray}
This gives
\begin{eqnarray}
 \frac{d r}{d t} \,\,\frac{d}{d r} \,L^{mech}_z (r,z)
 \ = \ - \,e \,r \,v_r \,B_z (r,z) .
\end{eqnarray}
Since $v_r = d r / d t$, we have
\begin{eqnarray}
 \frac{d}{d r} \,L^{mech}_z (r, z) \ = \ 
 - \,e \,r \,B_z (r,z) .
\end{eqnarray}
Integrating it from $r$ to $r = \infty$, we therefore get 
\begin{eqnarray}
 L^{mech}_z (r = \infty, z) \ - \ L^{mech}_z (r, z)
 \ = \ - \,e \,\int_r^\infty \,r^\prime \,
 B_z (r^\prime, z) \,d r^\prime .
\end{eqnarray}
Since  $L^{mech}_z (r = \infty, z) = m_0$ from the given initial condition, 
this becomes
 \begin{eqnarray}
 m_0 
 \ = \ L^{mech}_z (r, z) 
 \ - \ e \,\int_r^\infty \,r^\prime \,
 B_z (r^\prime, z) \,d r^\prime .
\end{eqnarray}
For finite but very large $r$ with $r \gg L$, 
we know that $A_\phi \sim 1 / r^2$, so that 
the Stokes theorem gives
\begin{eqnarray}
 2 \,\pi \,r \,A_\phi (r, z) &=& 
 \oint_{|\bm{x}| = r} \,\bm{A} (\bm{x}, z) \cdot d \bm{x}
 \ = \ \iint_{|\bm{x}| < r} \,\bm{B} \cdot \bm{n} \,d S \nonumber \\
 &=& - \,\iint_{|\bm{x}| > r} \,\bm{B} \cdot \bm{n} \,d S
 \ = \ - \,
 \int_{r}^{\infty}
 \,B_z (r^\prime, z) \,2 \,\pi \,r^\prime \,d r^\prime .
\end{eqnarray}
We thus arrive at the relation
\begin{eqnarray}
 m_0 \ = \ 
 L^{mech}_z (r, z) \ + \ e \,r \,A_\phi (r, z) . \label{OAM_finite_solenoid}
\end{eqnarray}
The 2nd term of the r.h.s. is nothing but the potential angular momentum
$L^{pot}_z = e \,(\bm{x} \times \bm{A}_\perp)_z = e \,r \,A_\phi$ in our
terminology. In fact, the above relation is again a special case of
our general relation
$L^{g.i.c.}_z \, = \, L^{mech}_z \, + \, L^{pot}_z$.
Eq.(\ref{OAM_finite_solenoid}) obtained above can be
interpreted as follows.
Although the electron mechanical OAM changes depending on the position, 
the change of the potential OAM compensates this change, such that the 
canonical OAM $m_0$ is a constant of motion throughout the whole space
outside the solenoid.

In going to quantum mechanics, the canonical OAM is quantized such that 
it takes integer values ($m = 0, \,\pm \,1, \, \pm \,2, \,\cdots \,$).                                
For large enough length of the solenoid, the eigen-function of the electron 
with the kinetic energy $k^2 / 2 \,M$ takes the following 
form \cite{GP1985}
\begin{eqnarray}
 \frac{1}{(2 \,\pi)^{1/2}} \,
 J_{|\alpha|} (k \,r) \,
 e^{\,i \,m \,\phi 
 \ - \ i \,\frac{k^2}{2 \,M} \,t}
 \hspace{10mm} (m = 0, \pm 1, \pm 2, \cdots) ,
\end{eqnarray}
where $J_\nu (z)$ is the standard Bessel function, while
\begin{eqnarray}
 \alpha \ = \ 
 m \ + \ e \,\int_r^\infty \,
 B_z (r^\prime, z) \,r^\prime \,d r^\prime .
\end{eqnarray}
For large enough length of solenoid, we also have
\begin{eqnarray}
 \int_r^\infty \,B_z (r^\prime,z) \,2 \,\pi \,r^\prime \,d r^\prime
 \ = \ - \,\Phi
 \hspace{10mm} (\mbox{\tt return flux}) ,
\end{eqnarray}
so that the quantity $\alpha$ approaches
\begin{eqnarray}
 \alpha \ \sim \ 
 m \ - \ 
 \frac{e}{2 \,\pi} \,\Phi \ = \ 
 m \ - \ 
 \beta ,
\end{eqnarray}
where $\beta \equiv \frac{e}{2 \,\pi} \,\Phi$, while the eigenfunction
approaches
\begin{eqnarray}
 \mbox{\tt eigen-function} \ \sim \ 
 \frac{1}{(2 \,\pi)^{1/2}} \,
 J_{\,|m - \beta|}  (k \,r) \,
 e^{\,i \,m \,
 \phi \,- \,i \,\frac{k^2}{2 \,M} \,t} .
\end{eqnarray}
Suppose now that we take the ultimate limit of $L \rightarrow \infty$.
In this limit, the return flux outside the solenoid disappear completely.
Curiously, even in this limit, the difference between the canonical
OAM and the mechanical OAM remains. In fact, in this limit,    
both of these as well as the potential OAM becomes constants, 
which are independent of spatial position of the electron. i.e.
\begin{equation}
 L^{mech}_z \ \stackrel{L \rightarrow \infty}{\longrightarrow} \ m - \ \beta, \ \ \ \ \ 
 L^{pot}_z \ \stackrel{L \rightarrow \infty}{\longrightarrow} \ \beta ,
\end{equation}
with $L^{g.i.c.}_z = L^{mech}_z + L^{pot}_z = m$ being unchanged.
This especially means that we could understand
the origin of non-zero potential angular momentum outside the
infinitely long solenoid as an action of
magnetic Lorentz force, even though the sign of the non-zero magnetic
field outside the solenoid is completely obliterated in the
limit $L \rightarrow \infty$.
The argument above illustrates fairly singular nature of an infinitely long 
solenoid prepared as an idealized setting of the AB-phase measurement.   
At the same time, however, it is thought to provide one nontrivial
explanation on the dynamical origin of non-zero orbital angular
momentum carried by the pure-gauge potential 
in the idealized Aharonov-Bohm solenoid problem.
In any case, we now realize that all these peculiar features in the
Aharonov-Bohm solenoid effect happen, just because
$\phi_{AB} \propto L_z^{pot} \propto r \,A_\phi$, while the behavior
of $A_\phi$ outside the infinitely-long solenoid is given by
$A_\phi \sim 1 / r$, so that $L^{pot}_z$ is {\it position-independent}
constant.

\subsection{Origin of potential angular momentum from general electromagnetic theory}

To resolve the seeming contradiction observed for the identity (\ref{L_pot})
in the case of the Aharonov-Bohm solenoid problem, more careful consideration of
surface terms is necessary. We can show that, with careful account
of surface terms, and by restricting the region of volume integral to the
outside region of the solenoid, the 2-dimensional version of (\ref{L_pot}) 
reads as
\begin{equation}
 \epsilon_0 \,\int_V \,d^2 x^\prime \,
 [ \bm{x}^\prime \times ( \bm{E}_\parallel (\bm{x}^\prime) \times
 \bm{B} (\bm{x}^\prime) ) ]_z \ = \ L^{pot}_z \ + \ S_1 \ + \ S_2 \ + \ S_3 ,
 \label{L_pot_origin}
\end{equation}
where
\begin{equation}
 L^{pot}_z \ = \ \int_V \,d^2 x^\prime \,\rho (\bm{x}^\prime) \,
 ( \bm{x}^\prime \times \bm{A}_\perp (\bm{x}^\prime) )_z \ = \ 
 e \,( \bm{x}_e \times \bm{A}_\perp (\bm{x}_e) )_z ,
\end{equation}
and
\begin{eqnarray}
 S_1 &=& \epsilon_0 \,\int_V \,d^2 x^\prime \,\nabla^{\prime j} \,
 \left[ (\nabla^{\prime j} A^0 (\bm{x}^\prime) ) \,
 (\bm{x}^\prime \times \bm{A}_\perp (\bm{x}^\prime) )_z \right] , 
 \label{S_1} \\
 S_2 &=& \epsilon_0 \,\int_V \,d^2 x^\prime \,\nabla^{\prime j} \,
 \left[ A^0 (\bm{x}^\prime) \,(\bm{x}^\prime \times \nabla^\prime)_z \,
 A^j_\perp (\bm{x}^\prime) \right] , \label{S_2} \\
 S_3 &=& \epsilon_0 \,\int_V \,d^2 x^\prime \,\left[ \nabla^\prime \times 
 ( A^0 (\bm{x}^\prime) \,\bm{A}_\perp (\bm{x}^\prime) ) \right]_z . \label{S_3}
\end{eqnarray}
Here, $\rho (\bm{x}^\prime) = e \,\delta (\bm{x}^\prime - \bm{x}_e)$
is the charge density by the electron. (Here, the position vector of
the electron is denoted as $\bm{x}_e$ instead of $\bm{x}$.)
The reason why we restrict the region of spatial integration to the
outside region of the solenoid is as follows.
Since the electron is not allowed to enter the inside region of the solenoid,
the Aharonov-Bohm solenoid effect is often treated as
a quantum mechanics of the electron in a doubly-connected space, 
i.e. a 2-dimensional domain from which the inside region of the solenoid
is excluded.

The longitudinal electric field $\bm{E}_\parallel$
in (\ref{L_pot_origin}) is thought to be generated by the charge of the electron. 
In principle, the motion of the electron may also generate the
magnetic field in the outside region of the solenoid. 
However, it can be discarded provided that
the velocity of the electron is much slower than the speed of light.

As already mentioned, the 1st term on the r.h.s. of  (\ref{L_pot_origin}) 
represents the potential angular momentum, while the other stand for
three surface integral terms, the explicit forms of which are given
by (\ref{S_1})-(\ref{S_3}). Since $\bm{x}^\prime$ in $S_1, S_2$ and $S_3$
is a variable of integration, we hereafter use the notation $\bm{x}$ to
avoid unnecessary notational complexity.
Although we do not need the explicit form of $A^0 (\bm{x})$ in the
following manipulation, we will show it for understandability.
It can easily be obtained as a solution of the Poisson equation : 
\begin{eqnarray}
 \Delta \,A^0 (\bm{x}) \ = \ - \,\frac{\rho (\bm{x})}{\epsilon_0} ,
\end{eqnarray}
with $\rho (\bm{x}) = e \,\delta (\bm{x} - \bm{x}_e)$. 
The solution of the above equation in the 2-dimensional space is given by
\begin{equation}
 A^0 (\bm{x}) \ = \ - \,\frac{e}{2 \,\pi \,\epsilon_0} \,
 \ln \left( | \bm{x} - \bm{x}_e | \,/\, r_0 \right) ,
\end{equation}
with $r_0$ being some constant with the dimension of length.

First, by using the divergence theorem of Gauss, $S_1$ can be
rewritten as a surface integral : 
\begin{equation}
 S_1 \ = \ - \,\int_{S_\infty} \,
 \left[ E^j_\parallel (\bm{x}) \,(\bm{x} \times \bm{A}_\perp (\bm{x}))_z \right]
 \cdot n^j \,d S 
 \ - \,\int_{S_R} \,
 \left[ E^j_\parallel (\bm{x}) \,(\bm{x} \times \bm{A}_\perp (\bm{x}))_z \right]
 \cdot ( - \,n^j) \,d S .
\end{equation}
Here, $S_\infty$ represents the circle with infinite radius $r_\infty$, whereas
$S_R$ does the circle with the radius $R$ of the solenoid.
The quantity $\bm{n}$ denotes the unit vector in the radial direction, i.e.
$\bm{n} = \bm{e}_r$.
With use of the relation $(\bm{x} \times \bm{A}_\perp)_z = \frac{1}{2} \,B_0 \,R^2$
together with $\bm{E}_\parallel = - \,\nabla \,A^0$,
we obtain
\begin{equation}
 S_1 \ = \ - \,\frac{1}{2} \,B_0 \,R^2 \,\epsilon_0 \,\int_{S_\infty} \,
 \bm{E}_\parallel (\bm{x}) \cdot \bm{e}_r \,d S \ + \ 
 \frac{1}{2} \,B_0 \,R^2 \,\epsilon_0 \,\int_{S_R} \,
 \bm{E}_\parallel (\bm{x}) \cdot \bm{e}_r \,d S .
\end{equation}
Since the source of the electric field $\bm{E}_\parallel$, which is just the electron,
is located in the region $R < |\bm{x}_e| < r_\infty$, the familiar Gauss law dictates
that
\begin{equation}
 \epsilon_0 \,\int_{S_\infty} \,\bm{E}_\parallel (\bm{x}) \cdot 
 \bm{e}_r \,d S \ = \ e, \hspace{5mm}
 \epsilon_0 \,\int_{S_R} \,\bm{E}_\parallel (\bm{x}) \cdot \bm{e}_r \,d S \ = \ 0.
\end{equation}
This gives
\begin{equation}
 S_1 \ = \ - \,e \,B_0 \,R^2 \ = \ - \,e \,(\bm{x} \times \bm{A}_\perp (\bm{x}))_z .
\end{equation}
We thus find that the surface term $S_1$ gives a contribution which precisely 
cancels the potential angular momentum term.

Next, we turn to the second surface term $S_2$. Utilizing again the integral
theorem of Gauss, we obtain
\begin{equation}
 S_2 \ = \ - \,\epsilon_0 \,
 \int_{S_\infty} \,\left[ A^0 \,(\bm{x} \times \nabla)_z \,A^j_\perp \right]
 \,n^j \,d S \ - \ \epsilon_0 \,
 \int_{S_R} \,\left[ A^0 \,(\bm{x} \times \nabla)_z \,A^j_\perp \right]
 (- \,\,n^j) \,d S .
\end{equation}
Now, using
\begin{equation}
 (\bm{x} \times \nabla)_z \,\bm{A}_\perp \ = \ 
 \frac{\partial}{\partial \phi} \,\left( \frac{1}{2} \,B_0 \,\frac{R^2}{r} \,\bm{e}_\phi \right)
 \ = \ - \,\frac{1}{2} \,B_0 \,\frac{R^2}{r} \,\bm{e}_r ,
\end{equation}
we find that
\begin{equation}
 S_2 \ = \ \frac{1}{2} \,B_0 \,R^2 \,\epsilon_0 \,\int_{S_\infty} \,A^0 \,d \phi \ - \ 
 \frac{1}{2} \,B_0 \,R^2 \,\epsilon_0 \,\int_{S_R} \,A^0 \,d \phi \,.
\end{equation}

Next, by using the Stokes theorem, $S_3$ can be rewritten as
\begin{equation}
 S_3 \ = \ - \, \epsilon_0 \,\int_{C_\infty} \,A^0 \,\bm{A}_\perp \cdot \bm{e}_\phi \,r \,d \phi
 \ + \ \epsilon_0 \,\int_{C_R} \,A^0 \,\bm{A}_\perp \cdot \bm{e}_\phi \,r \,d \phi .
\end{equation}
Here, $C_\infty$ and $C_R$ are the counterclockwise circle route with radius $r_\infty$ 
and $R$, respectively.
In our problem with 2-dimensional geometry, they can basically be identified 
with
$S_\infty$ and $S_R$. With use of $\bm{A}_\perp = \frac{1}{2} \,B_0 \,R^2 \,\bm{e}_\phi$,
the 3rd surface term $S_3$ then reduces to
\begin{equation}
 S_3 \ = \ - \,\frac{1}{2} \,B_0 \,R^2 \,\epsilon_0 \,\int_{S_\infty} \,A^0 \,d \phi \ + \ 
 \frac{1}{2} \,B_0 \,R^2 \,\epsilon_0 \,\int_{S_R} \,A^0 \,d \phi .
\end{equation}
One notices that the two surface terms $S_2$ and $S_3$ exactly cancel each other.
As a consequence, we finally arrive at the relation
\begin{equation}
 \epsilon_0 \,\int \,d^2 x^\prime \,\left[ \bm{x}^\prime \times 
 (\bm{E}_\parallel  (\bm{x}^\prime) \times \bm{B} (\bm{x}^\prime) ) \right]_z
 \ = \ L^{pot}_z \ + \ S_1 \ + \ S_2 \ + \ S_3 , \label{math_identity}
\end{equation}
with
\begin{equation}
 S_1 \ = \ - \,L^{pot}_z, \hspace{6mm} S_2 + \ S_3 \ = \ 0 .
\end{equation}
Since the l.h.s. of the above equation is apparently zero in the magnetic-field-free
region, the mathematical correctness of the identity (\ref{math_identity}) has been proved.
A crucial observation is that the surface term $S_1$ exactly cancels the potential
angular momentum term. This again exposes quite singular nature of the
infinitely long solenoid. This surface term may be thought
of as a reminiscence of the return magnetic flux, which is pushed away to the spatial
infinity in the limit of infinitely long solenoid.

\section{Summary and conclusion}
\label{Section:s4}

Coming back to the most fundamental question about the AB-effect, i.e. 
``Does it show the reality of the electromagnetic potential ?”, we
may answer as follows.
The Helmholtz theorem dictates that the transverse component of the 
vector potential is uniquely fixed from the static magnetic field distribution or
the stationary electric current distribution of the solenoid, 
so that it appears to be the most favorable candidate of
electromagnetic potential as
a “physical reality”. Probably, this is not far from the truth.
Nevertheless, it should still be kept in mind that the observable
in the AB-effect
is not the vector potential itself but its closed integral. 
This means that any other vector potentials obtained through gauge 
transformations can give the same answer to the AB-phase shift. 
It is this fact that still prevents us from concluding straightforwardly that
the transverse part or the "physical" component
of the vector potential is a direct observable of the AB-effect.
In either case, if we accept the fact that the vector potential outside the solenoid 
is generated by the magnetic field distribution (or current distribution) inside 
the solenoid, it appears to support the non-locality interpretation of the AB-effect. 
An important
point here is that it should not be interpreted as breaking of causality or
the action-through-medium principle, as we have argued.

A remarkable observation is that the vector potential outside the solenoid is 
a pure-gauge configuration, which is just consistent with the
absence of the magnetic field outside the solenoid.
Very curiously, however, it appears to carry some sort of orbital 
angular momentum.
According to Peshkin \cite{Peshkin1981}, the essence of the
AB-effect
is thought to be the quantization of the electron canonical OAM.
The canonical OAM (or its gauge-invariant version) in the AB-effect 
is not an observable, however.
The observable of the AB-effect is a fractional (non-integer) part of the 
mechanical OAM, which can be related to the potential OAM in our terminology.
One primitive question is why and how the pure-gauge potential outside the 
infinitely long solenoid can acquire nonzero orbital angular momentum.
We point out that there exist two totally different explanations 
based on two totally different local forces.

\begin{itemize}

\item Explanation based on the induced electric field by time-varying 
magnetic field.

\item Explanation based on the magnetic Lorentz force due to a
finite-length solenoid, and the subsequent operation of the
infinite-length limit.
\end{itemize}

The existence of these two totally different explanations based on two totally 
different local forces indicates a singular nature of the infinitely-long solenoid, 
which is thought to contain the essence of the AB-effect.

We have also tried to clarify the reason why the pure-gauge configuration
outside the infinitely long solenoid is able to carry non-zero orbital angular 
momentum on the basis of the general electromagnetic theory of angular momentum.
We showed that the key to resolve this seeming paradox is the unexpected role
of surface term, which precisely cancels the potential angular momentum. 
Undoubtedly, all the peculiarities of the Aharonov-Bohm solenoid setting of the
problem explained in the present paper appear to be the reason to allow various 
theoretical interpretations of the AB-effect, which look very different but
nevertheless most are correct in some sense.
The Aharonov-Bohm solenoid effect is still continuing to offer attractive 
challenge to wide area of theoretical physics.

\noindent
\section*{Acknowledgment}

\vspace{2mm}
\noindent
M.W. thanks the Institute of Modern Physics of the Chinese
Academy of Sciences in Lanzhou for hospitality.
Y.K. L.Z and P.-M.Z. are supported by the National Natural
Science Foundation of China (Grant No.11575254).
This work is partly supported by the Chinese Academy of Sciences
President's International Fellowship Initiative 
(No. 2018VMA0030 and No. 2018PM0028)







\vspace{8mm}
\noindent
\section*{References}


\end{document}